\documentclass[aps,prb,twocolumn,groupedaddress]{revtex4}
\usepackage{amsmath}
\usepackage{graphicx}
\begin{document}

\title{Cooling Dynamics of Photoexcited Carriers in Si Studied by Using Optical Pump and Terahertz Probe Spectroscopy}
\author{Takeshi Suzuki and Ryo Shimano}
\affiliation{Department of Physics, The University of Tokyo, Tokyo 113-0033, Japan}
\date{\today}

\begin{abstract}
We investigated the photoexcited carrier dynamics in Si by using optical pump and terahertz probe spectroscopy in an energy range between 2 meV and 25 meV. The formation dynamics of excitons from unbound e-h pairs was studied through the emergence of the 1s-2p transition of excitons at 12 meV (3 THz).  We revealed the thermalization mechanism of the photo-injected hot carriers (electrons and holes) in the low temperature lattice system by taking account of the interband and intraband scattering of carriers with acoustic and optical phonons.  The overall cooling rate of electrons and holes was numerically calculated on the basis of a microscopic analysis of the phonon scattering processes, and the results well account for the experimentally observed carrier cooling dynamics. The long formation time of excitons in Si after the above-gap photoexcitation is reasonably accounted for by the thermalization process of photoexcited carriers. 
\end{abstract}

\maketitle

\section{Introduction}
Carrier dynamics in photoexcited semiconductors has been a central issue in semiconductor physics for decades~\cite{shah1}. In particular, the carrier-phonon interactions have important effects in a wide range of optoelectronic devices, such as laser diodes and solar cells. Since the development of ultrafast lasers, numerous studies have been devoted to elucidating the photoexcited carrier dynamics by using nonlinear laser spectroscopic techniques such as optical pump and optical probe spectroscopy, four wave mixing, and transient gratings~\cite{shah1}. Time-resolved photo-luminescence has also been used to investigate hot carrier relaxation and exciton formation~\cite{leo1,leo2,szczytko1}. With the development of mid-infrared ultrafast light sources, the wavelength of the optical probe has been extended to the lower energy region and it has become possible to access the intraband transitions, i.e., the free carrier responses, or the intra-exciton transitions with sub-ps temporal resolution. As such, the mid-infrared probe method has been used to study, e.g., electron-hole (e-h) droplet formation in CuCl~\cite{nagai1} and conversion dynamics of orthoexcitons to paraexctions in Cu$_2$O~\cite{kubouchi}.

The advent of terahertz time-domain spectroscopy (THz-TDS) has enabled researchers to investigate the lower end of the terahertz frequency range ~\cite{nuss1, mittleman, tonouchi, exter, koch}. Unlike optical methods, THz-TDS enables the determination of the complex optical conductivity from which one can quantitatively evaluate the carrier density and the scattering time of the photoexcited carriers \cite{nuss2, greene, beard}, without resorting to the Kramers-Kronig relation. The technique has been used to investigate the buildup of plasma screening in a high-density e-h system in GaAs~\cite{huber1} and formation of excitons in GaAs quantum wells~\cite{huber2,kaindl1}.

Carrier dynamics in indirect gap semiconductors, in particular in Si, has also been extensively studied. Near-infrared and visible pump-probe studies~\cite{driel, sabbah, hase}, and more recently time-resolved photoemission spectroscopy~\cite{ichibayashi}, have revealed the ultrafast dynamics of hot carrier relaxation at very high carrier densities ($\gtrsim 10^{19}$ cm$^{-3}$). However, the photoexcited dynamics at lower carrier density in Si is not fully understood yet, despite their inevitable importance in optoelectronic applications. From the viewpoint of many-body quantum phenomena in e-h systems, an unambiguous understanding of the cooling dynamics of the photoexcited e-h system will be crucial for the realization of theoretically predicted low-temperature quantum degenerate phases such as exciton Bose-Einstein condensation and the e-h BCS state~\cite{griffin, kremp, ogawa}. A quantitative description on the slow formation time of excitons on the order of several hundreds of ps, as observed in our previous study~\cite{suzuki}, is also needed for gaining an understanding of the fundamental optical properties of Si.

The current study addressed the carrier thermalization dynamics of photoexcited e-h system in Si and the slow formation dynamics of excitons by using optical pump and THz probe measurements. We evaluated the time-dependent fraction of free carriers and excitons from the spectral shape analysis of the complex optical conductivity in the terahertz frequency range and estimated the effective carrier temperature by using the Saha equation. This analysis was partly motivated by the scheme reported by Kaindl et al.~\cite{kaindl1}. The cooling dynamics of the carrier system was analyzed numerically by calculating all the relevant phonon relaxation rates. The observed cooling rate was quantitatively reproduced without adjustable parameters. Accordingly, the observed long formation time of excitons in Si after above-gap photoexcitation was well accounted for by the carrier thermalization dynamics dominated by electron (hole) - phonon scattering.
\begin{figure*}[htpb]
	\begin{center}
		\includegraphics[width = 55 mm]{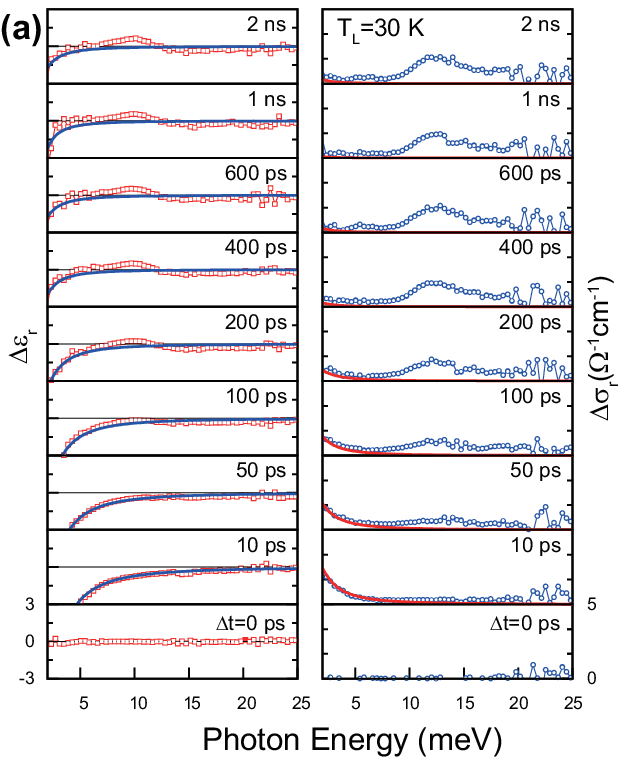}
		\includegraphics[width = 55 mm]{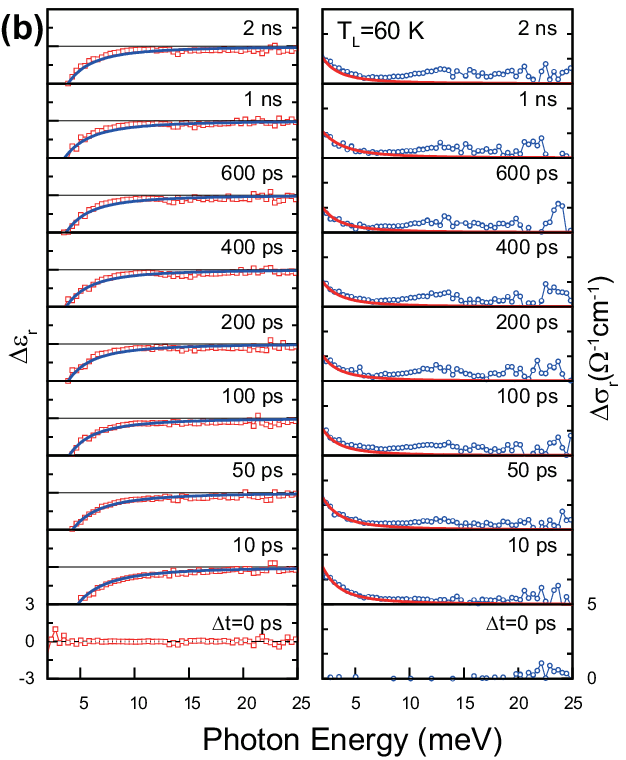}
\caption{(color online). Photo-induced change of dielectric function (left) and conductivity (right) at lattice temperature $T_L$ (a) 30 K and (b) 60 K for the excitation density of $I_p$=17 $\mu\text{J}/\text{cm}^{2}$ and for the indicated pump-probe delays $\Delta t$. The solid lines in each panel show the Drude model fitted to the low-energy region.}
\label{ex_form}
	\end{center}
\end{figure*}
\section{Experiments 
\label{exp}}
\subsection{Methods}
We used the optical pump and THz probe technique to study the transient photoexcited free carrier responses and to investigate the formation dynamics of excitons through the observation of the 1s-2p transition around 12 meV ($\sim$3 THz)~\cite{labrie}. The output from a multi-pass amplifier seeded by a 14-fs mode-locked Ti:sapphire oscillator was used as a light source.  The duration of the amplified pulse was 30 fs with the repetition rate of 1 kHz, and the center wavelength was 800 nm (1.55 eV).  The output from the amplifier was divided into three beams, respectively for THz generation, THz detection, and the optical pump.  The THz probe pulse was generated from a 300 $\mu$m thick GaP(110) crystal by optical rectification of the laser pulse.  The waveforms of the THz pulse transmitted after the sample with and without the optical pump were recorded by electro-optic sampling with a 300 $\mu$m GaP crystal~\cite{wu}.  The complex transmittance change was then obtained by Fourier transformation, thereby yielding a photo-induced change in the complex dielectric function, $\Delta \tilde{\epsilon}=\Delta\epsilon_1+i \Delta \sigma_1/\epsilon_0 \omega$, where $\Delta \sigma_1$ is the change in the real part of the optical conductivity.  The frequency range of our experimental setup was between 0.5 THz (2 meV) to 6 THz (25 meV).  To make a homogeneous optical excitation in the lateral direction, the sample was irradiated with a pump pulse with a spot size of 3 mm in diameter, which was larger than the THz-probe spot size of d=4$\lambda$ ($\lambda$ ranged from 0.05 mm (6 THz) to 0.6 mm (0.5 THz)) in our experimental setup. 

As a sample, we used a high purity and high resistivity (10 k$\Omega$ cm at room temperature) Si single crystal with an optically flat (001) surface.  To achieve a nearly homogeneous optical excitation in the depth direction, the sample was mechanically polished to a thickness of 22 $\mu$m, which was smaller than the penetration depth of 27 $\mu$m at 1.5 eV.  The sample surface was cleaned with hydrofluoric acid (HF) to remove the surface oxide layer.  The same results were also obtained without the HF treatment. The sample was freely mounted inside a cryostat.

\subsection{Formation dynamics of excitons 
\label{exciton_formation}}

\begin{figure}
	\begin{center}
		\includegraphics{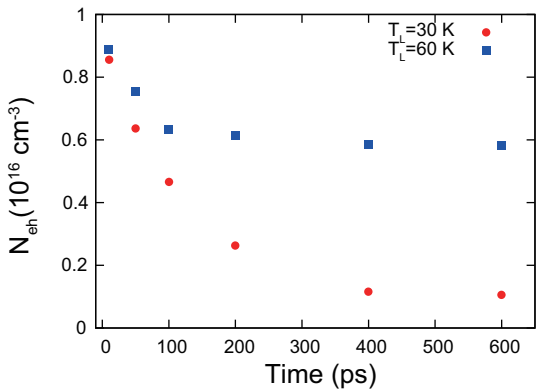}
\caption{(color online). The temporal change of the free e-h pair density $N_{eh}$ obtained from the fitting in Fig. \ref{ex_form}. }
		\label{density_dynamics}
	\end{center}
\end{figure}

We first studied the formation dynamics of excitons from the photoexcited unbound e-h pairs. The dots in Fig. \ref{ex_form} show the experimental results obtained at lattice temperatures of $T_L$=30 K (a) and 60 K (b). The excitation density was $I_p$=17 $\mu$J/cm$^2$ and the delay time $\Delta t$ between pump and probe pulses was varied.  From the excitation pulse energy flux and the absorption coefficient, the excited e-h pair density $N$ can be estimated to be (1.3 $\pm$ 0.4) $\times$ 10$^{16}$ cm$^{-3}$. Please note that this value is below the Mott density at 30 K, $N_{\text{Mott}}=7.4 \times 10^{16} \text{ cm}^{-3}$, calculated from the random phase approximation~\cite{norris}. The dielectric function and the conductivity spectra 10 ps after the photoexcitation are well reproduced by the Drude model, according to which the dielectric function is expressed as
\begin{equation}
\epsilon(\omega) = \epsilon_b - \frac{N_{eh} e^2}{\epsilon_0 m^{\ast}} \frac{1}{\omega(\omega + i\gamma)} \label{drude},
\end{equation}
where $\epsilon_b=11.7$ is the background dielectric constant, $\gamma$ the damping constant, $N_{eh}$ the free e-h pair density, $\epsilon_0$ the vacuum permittivity and $m^{\ast}$=0.16$m_0$ the optical mass of free carriers in Si~\cite{riffe}, respectively. The Drude component decreases with increasing delay time,and a peak emerges around 10-12 meV, an energy that corresponds to the 1s-2p transition of excitons in Si. To estimate the free e-h pair density at each delay time, we can fit the data with the Drude model in the low energy region (2.0-7.8 meV), where the Drude component dominates the spectrum. The solid lines in Fig. \ref{ex_form} plot the fitting curves, without the contribution from the exciton component. Figure \ref{density_dynamics} shows the temporal change of $N_{eh}$ obtained by the Drude fitting. At $T_L=30 \text{ K}$, $N_{eh}$ continues to decrease until about 400 ps. Contrary to this free carrier behavior, the exciton density, as indicated by the 1s-2p absorption at 12 meV, gradually increases after the photoexcitation and reaches a constant value after 400 ps (Fig. \ref{ex_form} (a)). Accordingly, the temporal decrease in $N_{eh}$ can be attributed to the formation of excitons. It should be noted here that the lifetime of the indirect excitons in Si is 1 $\mu$s~\cite{hammond}, so that the dynamics observed in the current temporal region is free from the lifetime effect. Since it is considered that thermalization of the electron system occurs within 1 ps through the electron-electron collisions~\cite{camescasse}, the observed long formation time of excitons can be presumably attributed to the thermal relaxation process of carriers in the lattice system. As shown in Fig. \ref{density_dynamics}, a shorter decay time of about 100 ps was observed at $T_L=60 \text{ K}$; i.e., the thermalization time of the carrier-lattice system is shorter than at $T_L = 30 \text{ K}$.

\subsection{Thermal ionization of excitons 
\label{temp_dep}}
\begin{figure}
	\begin{center}
		\includegraphics{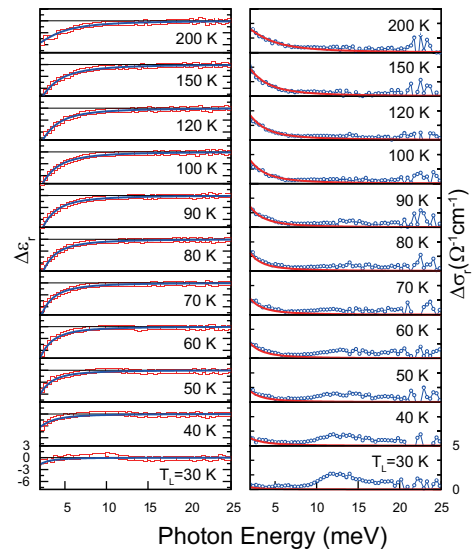}
\caption{(color online). Photo-induced change in dielectric function (left) and conductivity (right) for the excitation density of 17 $\mu\text{J}/\text{cm}^{2}$ at various lattice temperatures. The solid lines in each panel show the Drude model fitted to the low-energy region.}
		\label{fig_temp}
	\end{center}
\end{figure}

\begin{figure}[htbp]
	\begin{center}
		\includegraphics{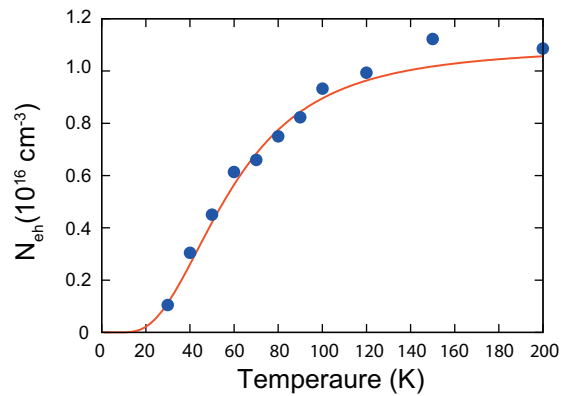}
\caption{(color online). The lattice temperature dependence of the free carrier density $N_{eh}$. The dots show the experimentally obtained $N_{eh}$. The solid line shows the prediction of the Saha equation.}
		\label{drude_data_cal}
	\end{center}
\end{figure}
Since the exciton formation is complete at 400 ps at $T_L\geq 30$ K, we can see that excitons and phonons should reach thermal equilibrium after 400 ps. To confirm the thermal equilibrium condition, we set the delay time to $\Delta t = 2 \text{ ns}$ and investigated the stability of excitons against the thermal ionization by varying the lattice temperature (Fig. \ref{fig_temp}). The excitation density was the same as for the results in Fig. \ref{ex_form}, $I_p$=17 $\mu$J/cm$^2$. With increasing $T_L$, the spectrum clearly changes from a Lorentz-like shape (excitonic 1s-2p intraband transition) to a Drude-like one (free carrier), indicating exciton ionization due to thermal activation. To estimate the free e-h pair density quantitatively, we performed a fitting of the Drude model following the procedure described in the previous section. The dots in Fig. \ref{drude_data_cal} are plots of the resulting $N_{eh}$ as a function of $T_{L}$. 
Now, let us compare our experimental results with the predictions of the Saha equation~\cite{kaindl1}. The Saha equation describes the ratio of the density of excitons $N_{X}$ and free carriers after statistical equilibration of their chemical potential in the Boltzmann limit. For a 3D case, it is
\begin{equation}
	\frac{ N_{eh}^2 }{ N_X} 
	= \frac{1}{4} \left( \frac{2\mu k_B T_C}{\hbar \pi} \right) ^{3/2}
	\exp \left( - E_0/k_B T_C \right),
	\label{saha_eq}
\end{equation}
where $\mu$=0.123$m_0$ and $E_0$=14.4 meV are the exciton reduced mass~\cite{lipari} and the binding energy~\cite{shaklee}. For a given total e-h pair density $N=N_{eh}+N_X$, Eq. (\ref{saha_eq}) yields the relationship between carrier temperature $T_C$ and free carrier density $N_{eh}$. The solid line in Fig. \ref{drude_data_cal} is the $N_{eh}$ obtained from Eq. (\ref{saha_eq}) with a fixed density $N$=1.1$\times$10$^{16}$ cm$^{-3}$ (as estimated from the excitation pulse energy flux) and it shows good agreement with the experimentally obtained value of $N_{eh}$. This agreement ensures the validity of our Drude model analysis and also the assumption that our experimental condition is at the Boltzmann limit (high lattice temperature and low excited e-h density).

\subsection{Cooling dynamics 
\label{cooling_dynamics}}
\begin{figure}
	\begin{center}
		\includegraphics{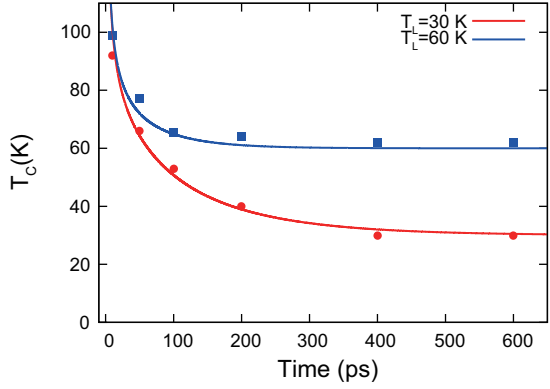}
\caption{(color online). The temporal change in carrier temperature $T_C(t)$. The dots show the experimental data obtained from the results in Fig. \ref{density_dynamics} and the Saha equation. The solid lines show the calculated thermodynamics considering the carrier-phonon interaction. }
		\label{thermodynamics}
	\end{center}
\end{figure}

Considering that the free carriers and excitons can reach equilibrium within the experimental temporal resolution of THz-TDS, $\sim $ 2 ps, we can use Saha equation to deduce the transient carrier temperature $T_C(t)$ from the transient free carrier density $N_{eh}(t)$. The dots in Fig. \ref{thermodynamics} represent $T_C(t)$, converted from Fig. \ref{density_dynamics} using the Saha equation. The plot for the higher lattice temperature $T_L$=60 K shows a fast thermal relaxation of carriers ($\sim$ 100 ps), while the plot for the lower temperature shows a relatively slow one ($\sim$ 400 ps). The reason for this difference can be attributed to the phonon types involved in the thermodynamics, as will be discussed in the following section. 

\section{Analysis of the cooling mechanisms 
\label{analysis}}
To describe the cooling mechanism of the electron system from the microscopic point of view, we calculate the thermal relaxation dynamics considering the interactions between carriers (electrons, heavy-holes and light-holes) and phonons.

\subsection{Electrons}
\begin{figure}[t]
	\begin{center}
		\includegraphics{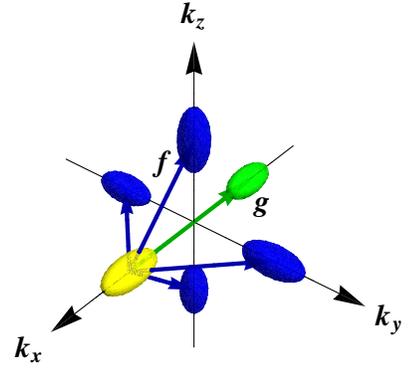}
\caption{(color online). The interband scattering processes in Si. $g$ scattering means scattering to the opposite band, and $f$ scattering means a transition to one of the remaining four bands.}
		\label{interband_scat}
	\end{center}
\end{figure}

For electrons, we consider the scattering process of intraband acoustic phonons and interband phonons. The energy-loss rate per electron in a Maxwellian distribution of temperature $T_e$ is given by~\cite{conwell}
\begin{align}
\left < \frac{d E_e}{dt} \right >_{\text{ac}} =
&- \frac{8(2)^{1/2}}{\pi^{2/3}} \frac{\Xi_0^2 m_{t}^{2} m_{l}^{1/2} }{\hbar ^4 \rho}  \nonumber \\
& \times ( k_B T_e)^{3/2}\left(1 - \frac{T_L}{T_e} \right),
\label{electron_ac}
\end{align}

\begin{align}
\left < \frac{dE_e}{dt} \right >_{\text{inter}}=
&-\left( \frac{2}{\pi} \right)^{1/2} \sum_{i} \frac{Z_i \left( D_t K \right)_{i}^2 m_t m_l^{1/2} }{\pi \hbar^2 \rho} \nonumber \\ 
&\times  \left( k_{\text{B}} T_e \right)^{1/2} \mathcal{B}(x^{i}_{0},x^{i}_{e}).
\label{electron_inter}
\end{align}
The subscript ``ac" and ``inter" denote the intraband acoustic deformation potential and interband scattering, respectively. $m_l$ and $m_t$ are the longitudinal and transverse effective masses of the electrons~\cite{riffe},  $\Xi_0$ is the intraband acoustic deformation potential, and $\left(D_t K \right)_i$ is the interband deformation potential involving the phonon type $i$, the density of the crystal $\rho$=2.33 g/cm$^3$, and the number of possible final equivalent valleys $Z_i$ = 1 ($g$ scattering) or 4 ($f$ scattering) (see Fig. \ref{interband_scat}). $\mathcal{B}(x^{i}_{0},x^{i}_{e})$ is 
\begin{gather}
\mathcal{B}(x^{i}_{0},x^{i}_{e}) = \frac{e^{(x^{i}_{0} - x^{i}_{e} )} - 1}{e^{x^{i}_{0}} - 1} \frac{x^{i}_{e}}{2} e^{x^{i}_{e}/2} K_1 \left( \frac{x^{i}_{e}}{2} \right), \\
x^{i}_{0} = \frac{\hbar \omega_i}{k_{\text{B}} T_L}, \\
x^{i}_{e} = \frac{\hbar \omega_i}{k_{\text{B}} T_e},
\end{gather}
where $\omega_i$ is the phonon frequency for the phonon type $i$, and $K_1$ is a Bessel function of the second kind with imaginary argument. Table \ref{electron-parameter} lists the physical parameters used in the calculation.

\begin{figure}[b]
	\begin{center}
		\includegraphics{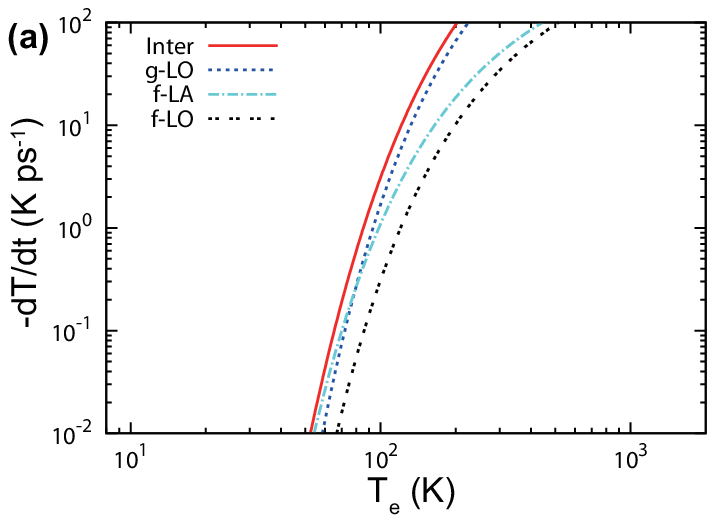}
		\includegraphics{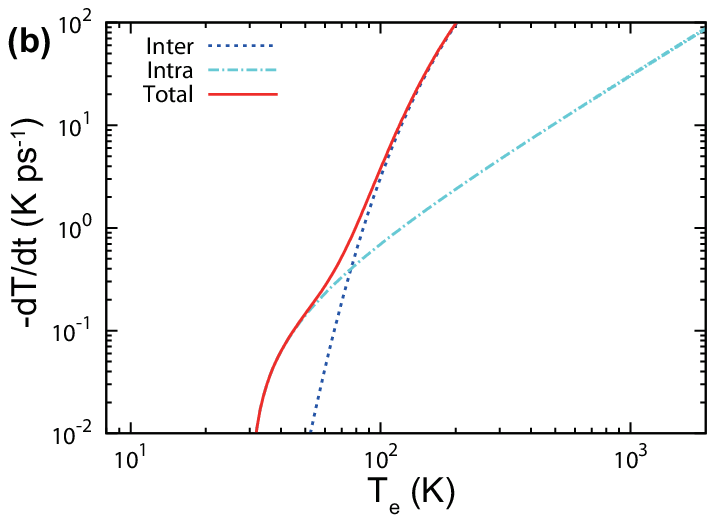}
\caption{(color online). Calculated cooling rate per electron, $-dT_e/dt$, for a thermal (Maxwell-Boltzmann) distribution with electron temperature $T_e$. The lattice temperature $T_L$ is fixed to 30 K. Cooling rates due to different interband scatterings are shown in (a), where each line denotes the following processes; g-LO: $g$ scattering with longitudinal optical phonons, f-LA: $f$ scattering with longitudinal acoustic phonons, f-LO: $f$ scattering with longitudinal optical phonons, and Inter: the sum of above three contributions. A comparison of cooling rates between interband scattering and intraband acoustic phonon scattering is shown in (b), where each line denotes the following contribution; Inter: the sum of interband scatterings (corresponding to Inter in (a)), Intra: intraband deformation scattering by acoustic phonons, and Total: the sum of all scatterings.}
		\label{cooling_rate_ele_30k}
	\end{center}
\end{figure}

\begingroup
\squeezetable
\begin{table}
	\begin{center}
		\caption{Band parameters of electrons~\cite{jacoboni}}
		\label{electron-parameter}
		\begin{ruledtabular}
		\begin{tabular}{ccc}
		 &Value & Units \\ \hline
		$m_l$ & 0.9163 & $m_0$\\
		$m_t$ & 0.1905 & $m_0$\\
		$\Xi_0$ & 9.0 & eV \\
		$(\hbar \omega)_{g-\text{LO}}$ & 720 & K \\
		$(D_t K)_{g-\text{LO}}$ & 11.0 & 10$^{8}$eV/cm \\
		$(\hbar \omega)_{f-\text{LA}}$ & 550 & K \\
		$(D_t K)_{f-\text{LA}}$ & 2.0 & 10$^{8}$eV/cm \\
		$(\hbar \omega)_{f-\text{TO}}$ & 685 & K \\
		$(D_t K)_{f-\text{TO}}$ & 2.0 & 10$^{8}$eV/cm \\
		\end{tabular}
		\end{ruledtabular}
	\end{center}
\end{table}
\endgroup

Because we consider the distribution function in the Boltzmann limit, $T_e$ is related to the average energy according to the equipartition theorem:
\begin{equation}
\left<E_e \right> = \frac{3}{2} k_B T_e.
\label{equipartition-law}
\end{equation}
Using Eq. (\ref{equipartition-law}), we can calculate the cooling rate $\left< dT_e/dt \right>$. Figure \ref{cooling_rate_ele_30k} (a) shows the cooling rate per electron due to interband scattering, and Fig. \ref{cooling_rate_ele_30k} (b) shows a comparison of coolings due to intraband scattering and interband scatterings.  As the phonon frequency increases, its contribution to the cooling rate becomes larger. In the present case, $g$ scattering with LO phonons contributes the most to the cooling process. Figure \ref{cooling_rate_ele_30k} (b) indicates that the contribution from the intraband scattering becomes larger than the interband scattering below 70 K, and the total cooling rate becomes smaller as $T_e$ decreases.

The present calculation does not include $g$ scattering by TA and LA phonons or $f$ scattering by TA phonons. Although these scatterings are forbidden in the selection rules~\cite{streitwolf, lax}, there are several experimental indications~\cite{canali} that they are active in electron transport phenomena in Si. These claims are based on the fact that the selection rules are evaluated for electron transitions between points exactly on the $\left<100\right>$ axes, while, in general, electrons undergo transitions between points relatively far from these axes. In the present case, however, the distribution function of electrons is not considered to be far from these axes because the density and temperature are relatively low, so we did not include these scatterings in our calculations.

\subsection{Holes}
Next, we consider phonon-scattering of holes. Because of the complexity of the valence band structure with its peculiarities of double degeneracy (heavy- and light-hole bands), nonparabolicity and anisotropy, the full description of the transitions could be quite complicated. However, for the present experiment, we can simplify the model as follows. Firstly, we consider only the intraband transitions for holes. Secondly, we include the acoustic-deformation (ac) and the nonpolar-optical-deformation (nop) potential as the scattering potentials. Thirdly, we replace the nonparabolic and anisotropic band structures with parabolic and isotropic ones with effective masses $m_i$ ($i$=light hole (lh), heavy hole (hh)). (The validity of this simplification will be discussed later.) Accordingly, the energy-loss rates per lh and hh in a Maxwellian distribution of temperature $T_{i}$ ($i$=lh, hh) are~\cite{conwell}
\begin{align}
\left < \frac{d E_i}{dt} \right >_{\text{ac}} =
&- \frac{8(2)^{1/2}}{\pi^{2/3}} \frac{E_1^2 m_{i}^{5/2} }{\hbar ^4 \rho}  \nonumber \\
& \times ( k_B T_i)^{3/2}\left(1 - \frac{T_L}{T_i} \right),
\label{hole_ac}
\end{align}

\begin{align}
\left < \frac{dE_i}{dt} \right >_{\text{nop}}=
&-\left( \frac{2}{\pi} \right)^{1/2} \frac{\left( D_t K \right)^2 m_i^{3/2} }{\pi \hbar^2 \rho} \nonumber \\ 
&\times  \left( k_{\text{B}} T_i \right)^{1/2} \mathcal{B}(x_{0},x_{i}),
\label{hole_nop}
\end{align}
for each scattering potential, where $E_1$ and $(D_tK)$ are the acoustic- and the optical-deformation potential, respectively. Table \ref{hole-parameter} lists the physical parameters used in the calculation.

\begingroup
\squeezetable
\begin{table}[h]
	\begin{center}
		\caption{Band parameters of holes~\cite{jacoboni}}
		\label{hole-parameter}
		\begin{ruledtabular}
		\begin{tabular}{ccc}
		 &Value & Units \\ \hline
		$m_{\text{lh}}$ & 0.154 & $m_0$\\
		$m_{\text{hh}}$ & 0.523 & $m_0$\\
		$E_1$ & 5.0 & eV \\
		$(\hbar \omega)_{\text{nop}}$ & 735 & K \\
		$(D_t K)$ & 6.0~\cite{jacoboni2} & 10$^{8}$eV/cm \\
		\end{tabular}
		\end{ruledtabular}
	\end{center}
\end{table}
\endgroup

\begin{figure}
	\begin{center}
		\includegraphics{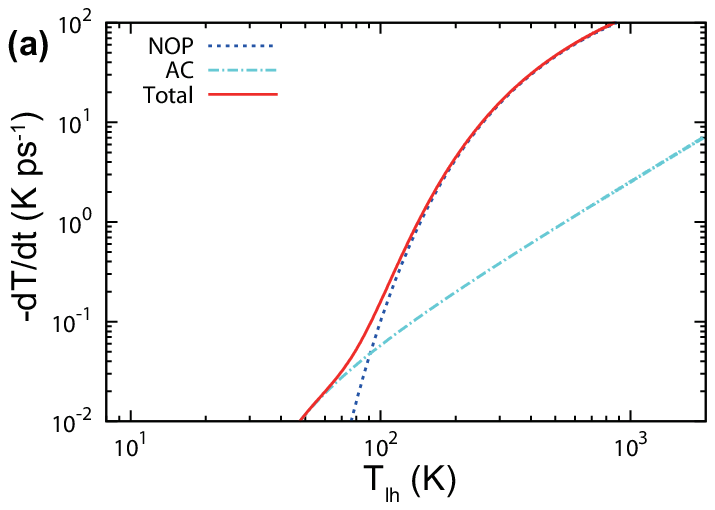}
		\includegraphics{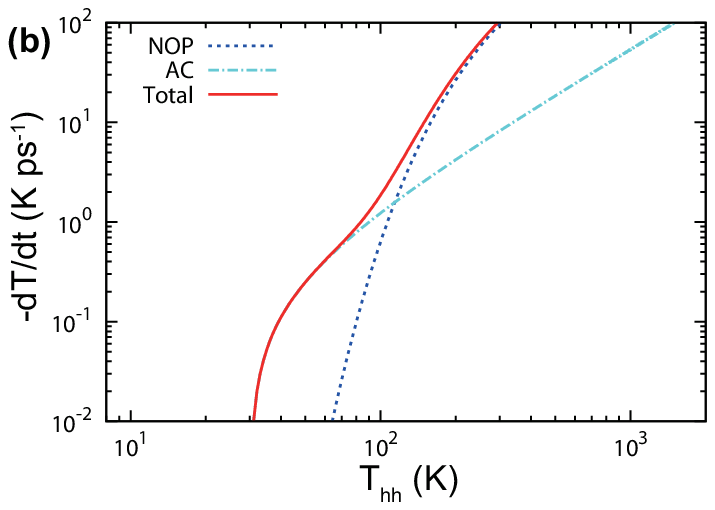}
\caption{(color online). Calculated cooling rate per light hole (a) and heavy hole (b), $-dT_i/dt$, for a thermal (Maxwell-Boltzmann) distribution with light and heavy hole temperatures $T_i$. The lattice temperature $T_L$ is fixed to 30 K. Each line denotes the following scatterings; NOP: non-polar-optical phonon scattering, AC: acoustic phonon scattering, Total: the sum of these two scatterings.}
		\label{cooling_rate_hole_30k}
	\end{center}
\end{figure}

By applying Eq. (\ref{equipartition-law}) to the case of holes, we can also calculate the cooling rate $\left< dT_i/dt \right>$. Figure \ref{cooling_rate_hole_30k} shows the cooling rate per light hole (a) and heavy hole (b) in the form of a comparison of intraband and interband scattering. The cooling rates behave similarly, as in the case of electrons. By comparing Fig. \ref{cooling_rate_hole_30k} (a) and (b), one can see that the heavy holes cool faster than light holes. This is because the cooling rate depends on the mass (Eq. (\ref{hole_ac}),(\ref{hole_nop})). 

Now let us discuss the validity of the simplified model. First, the nonparabolicity and the anisotropy of the valence bands should be taken into account in the high density or high temperature condition because the deviation from the spherical and parabolic bands becomes pronounced when holes reach energies far from the bottom of the valence bands. Under our conditions, however, the density is not so high such that the chemical potential for holes is $-9$ meV with a carrier density of $n=1\times10^{16}$ cm$^{-3}$ at 30 K. The thermal energy is $k_{\text{B}}T\sim3$ meV, and thus the energy of holes $\epsilon_\text{h}$ is distributed almost at the bottom of the valence bands compared to the spin orbit energy $\epsilon_{\text{so}}$ ($\epsilon_{\text{so}}$=44 meV in Si). Therefore, the condition $\epsilon_{\text{h}}\ll\epsilon_{\text{so}}$ is fulfilled, and the valence bands can be approximated to be spherical and parabolic. Next, the treatment of interband and intraband transition in holes is accomplished by considering the overlap factor $G(\nu)$, which is given by~\cite{combescot}
\begin{equation}
G(\nu) =
	\begin{cases}
	\frac{1}{4} \left( 1+ 3 \cos^2 \nu \right) : \text{intraband transition} \\
	\frac{3}{4} \sin^2 \nu : \text{interband transition},
	\end{cases}
\end{equation}
where $\nu$ is the scattering angle. The interband transition that corresponds to the large-angle scattering can be neglected from the following reason. The large-angle scattering requires the participation of phonons having large wave number. Due to the linear dispersion relation, these acoustic phonons with large wave number have high energy. When $T_L$ becomes low, the number of such high energy acoustic phonons becomes small. Therefore, in our experimental condition (relatively small $T_L$), the small-angle scattering is considered as the dominant process in the cooling mechanism.

\subsection{Cooling dynamics}
Having calculated each phonon-scattering process’s contribution to the cooling rate, we can now calculate the thermodynamics. For this, we used an algorithm that was developed to describe the photoexcited carrier dynamics in GaAs~\cite{leo1}. The algorithm proceeds as follows:  In the first step, we determine the initial carrier temperature $T_C(0)$ from the excitation photon energy $h \nu = 1.55$ eV, band gap $E_g$ = 1.17 eV, and the involved phonon energy for the indirect optical transition in Si, $\hbar \omega$=0.05 eV:
\begin{equation}
k_B T_C(0) = \frac{h \nu - E_g - \hbar \omega} {2}.
\label{initial-temperature}
\end{equation}
The resultant initial carrier temperature $T_C(0)$ is 2000 K. In the second step, we calculate each temperature, i.e., for electron ($T_e$) and holes ($T_i$ with $i$=hh,lh), after a finite time interval ($\Delta t$):
\begin{align}
\frac{3}{2} k_B T_{e} (t &+ \Delta t) = \frac{3}{2} k_B T_{C} (t) \nonumber \\
&+ \left( \left < \frac{d E_e}{dt} \right >_{\text{ac}} + \left < \frac{dE_e}{dt} \right >_{\text{inter}} \right) \Delta t,
\label{second-step-electron}
\end{align}

\begin{align}
\frac{3}{2} k_B T_{i} (t &+ \Delta t) = \frac{3}{2} k_B T_{C} (t) \nonumber \\
&+ \left( \left < \frac{d E_i}{dt} \right >_{\text{ac}} + \left < \frac{dE_i}{dt} \right >_{\text{nop}} \right) \Delta t
\label{second-step-hole}
\end{align}
In our calculation, $\Delta t$ is fixed to be 0.05 ps. In the third step, we unify the temperatures into one carrier temperature. The temperatures after $\Delta t$ are different because of the difference of energy loss rates of electrons, light holes, and heavy holes. By taking into account that the scattering between carriers is much faster than the scattering between carriers and phonons, we must unify the temperatures at each step. The total energy density in each system is obtained by the average energy multiplied by its density ($\left< E\right> \times n$). Considering that the average energy (Eq. (\ref{equipartition-law})) is proportional to the temperature, and that energy conserved in the carrier scatterings, the carrier temperature is given by the following equation: 
\begin{equation}
T_C(t) = \frac{N_e T_e(t) + N_{\text{lh}} T_{\text{lh}}(t) + N_{\text{hh}} T_{\text{hh}}(t)}{N_e + N_{\text{lh}} + N_{\text{hh}}}.
\label{unify}
\end{equation}
The densities of the systems are related as follows:
\begin{gather}
N_e = N_{\text{lh}} + N_{\text{hh}}, \label{density-relation-ele-hole} \\
\frac{N_{\text{lh}}}{N_{\text{hh}}} = \left( \frac{m_{\text{lh}}}{m_{\text{hh}}}\right)^{3/2}. \label{density-relation-hole}
\end{gather}
Equation (\ref{density-relation-ele-hole}) is true from the charge neutrality condition, and Eq. (\ref{density-relation-hole}) is true by virtue of that in the Boltzmann distribution, the density of states of ideal gas particles having mass $m$ in a 3D system is proportional to $m^{3/2}$. Thee fourth step returns to the second step with the new input parameter $T_C(t+\Delta t)$ and the steps are repeated.

The calculated results are shown by solid lines in Fig. \ref{thermodynamics}, and they are in very good agreement with the experimental data. Please note that no adjustable parameters were used. The rapid drop in temperature is due to interband scattering of electrons and optical phonon scattering of holes. The subsequent slow decline in temperature is due to the intraband phonon-scattering of electrons and holes by acoustic phonons. The calculated cooling rates in Fig. \ref{cooling_rate_ele_30k} and Fig. \ref{cooling_rate_hole_30k} indicate that the cooling time of the carrier system becomes longer at lower $T_L$.

Finally, let us address the scattering processes neglected by our calculation. Scattering from impurities would give only a small contribution because we used high purity Si, from which only free excitons were observed in the terahertz spectra. The screening effect of electron-phonon interactions by the electron gas is known to be important in the high density regime~\cite{yoffa}, above $N_{eh}=10^{19}$ cm$^{-3}$. In our case, however, the density is much smaller, and thus we believe that the screening effects are negligible. The simplified treatment using Maxwell-Boltzmann distribution function and also neglecting lattice heating may become inaccurate at very low $T_L$. Regarding this possibility, a more sophisticated calculation remains as future work.

\section{Summary}
We have investigated the cooling dynamics of photoexcited carriers in Si by broadband terahertz time-domain spectroscopy.  The formation dynamics of indirect excitons was revealed through the observation of the 1s-2p transition of excitons around 3 THz. The relatively long formation time of excitons ranging from 100 ps to several hundreds of ps dependent on the lattice temperature is attributed to the thermal relaxation time of the carriers in the lattice system. To view the cooling dynamics quantitatively, we determined the carrier temperature as follows. First, we evaluated the fraction of free carriers and excitons from the spectral shape analysis of the complex optical conductivity. From the obtained ratio of the free carrier density to the total e-h pair density, we determined the transient carrier temperature by using the Saha equation. To describe the observed cooling dynamics, we numerically calculated the carrier-phonon relaxation rate by taking into account all the relevant electron(hole)-phonon interactions in Si. We found that, at high temperatures, the carrier cooling process is dominated by the interband scattering of electrons with optical and acoustic phonons and by the intraband scattering of holes with optical phonons. At low temperatures (below $\sim$100 K), on the other hand, the contribution from the intraband acoustic phonons becomes dominant both for electrons and holes. The overall cooling rate of the e-h system was quantitatively calculated, and the calculated rate was in good agreement with the experimental result without adjustable parameters. This precise understanding of the electron-phonon interactions and the cooling mechanism of photoexcited e-h system in Si would be valuable for optoelectronics and photonics applications ~\cite{hochberg}, as well as for the realization of theoretically predicted low temperature quantum degenerate phases of e-h systems~\cite{griffin}. 
This work was partially supported by grants-in-aid (No. 22244036, No. 21104505), from MEXT, Japan

\end{document}